\documentclass[useAMS,usenatbib,usegraphicx]{mn2e}

\usepackage{times,epsfig}

\def\gs{\mathrel{\raise0.35ex\hbox{$\scriptstyle >$}\kern-0.6em
\lower0.40ex\hbox{{$\scriptstyle \sim$}}}}
\def\ls{\mathrel{\raise0.35ex\hbox{$\scriptstyle <$}\kern-0.6em
\lower0.40ex\hbox{{$\scriptstyle \sim$}}}}
\def\m@th{\mathsurround=0pt }
\def\eqalign#1{\null\,\vcenter{\openup1\jot \m@th
 \ialign{\strut\hfil$\displaystyle{##}$&$\displaystyle{{}##}$\hfil
 \crcr#1\crcr}}\,}

\title[Submm environments of $z>5$ quasars]
      {Environments of $z>5$ quasars: searching for protoclusters
       at submillimetre wavelengths}
\author[Priddey et al.]
       {R.\,S.\ Priddey,$^1$ R.\,J.\ Ivison$^{2,3}$ and K.\,G.\ Isaak$^4$
        \vspace*{1mm}\\
        $^1$ Centre for Astrophysics Research, Science and Technology Research
             Centre, University of Hertfordshire, College Lane, Herts AL10 9AB\\
        $^2$ UK Astronomy Technology Centre, Royal Observatory, Blackford Hill,
             Edinburgh EH9 3HJ\\
        $^3$ Institute for Astronomy, University of Edinburgh, Blackford Hill,
             Edinburgh EH9 3HJ\\
        $^4$ School of Physics \& Astronomy, Cardiff University, The Parade,
             Cardiff CF24 3AA
}

\date{\fbox{\sc Draft dated: \today\ }}
\pagerange{\pageref{firstpage}--\pageref{lastpage}}
\pubyear{2007}

\begin{document}

\maketitle

\label{firstpage}

\begin{abstract}
The most massive haloes at high redshift are expected, according to
hierarchical cosmologies, to reside in the most biased density fields.
If powerful active galactic nuclei (AGN) are expected to exist anywhere in the
early Universe ($z>\rm 5$), it is within these massive haloes. The most
luminous of these AGN, powered by supermassive black holes (SMBHs)
$\sim$10$^9$\,M$_{\odot}$, thereby present an opportunity to test models of
galaxy formation. Here, we present submillimetre (submm) continuum
images of the fields of three luminous quasars at $z>\rm 5$, obtained
at 850 and 450\,$\mu$m using the Submm Common-User Bolometer Array
(SCUBA) on the James Clerk Maxwell Telescope (JCMT). N-body
simulations predict that such quasars evolve to become the central
dominant galaxies of massive clusters at $z=\rm 0$, but at $z=\rm 5-6$
they are actively forming stars and surrounded by a rich
proto-filamentary structure of young galaxies. Our purpose in taking
these images was to search for other luminous, star-forming galaxies
in the vicinity of the signpost AGN and thus associated with such a
protocluster. Two of the quasar host galaxies are luminous SMGs in
their own right, implying star-formation rates (SFRs)
$\sim$10$^3$\,M$_{\odot}$\,yr$^{-1}$.  Despite the coarse 850-$\mu$m
beam of the JCMT, our images show evidence of extended emission on a
scale of $\sim$100-kpc from at least one quasar -- indicative of a
partially resolved merger or a colossal host galaxy. In addition, at
$>$3$\sigma$ significance we detect 12 (5) submm galaxies (SMGs) at
850\,$\mu$m (450\,$\mu$m) in the surrounding fields. Number counts
of these SMGs are comparable with those detected in the fields of
$z\sim\rm 4$ radio galaxies, and both samples are systematically overabundant
relative to blank-field submm surveys. Whilst the redshift-sensitive
850\,$\mu$m/450\,$\mu$m and 850\,$\mu$m/1.4\,GHz flux density ratios
indicate that some of these SMGs are likely foreground objects, 
the counts suggest that many probably lie in the same large-scale
structures as the quasars.
\end{abstract}

\begin{keywords}
\end{keywords}

\section{Introduction}

The search for $z>\rm 5$ galaxies has taken on a new urgency in recent
years. The epochal discovery of \citet{gp65} absorption in the spectra
of $z>\rm 6$ quasars \citep{becker01} suggests that reionisation was
complete by $z=\rm 6$. On the other hand, polarisation results from
the {\it Wilkinson Microwave Anisotropy Probe} ({\it WMAP}) point to
the onset to reionisation at $z\sim\rm 12$ \citep{spergel07}.  A
primary goal of observational cosmology is, consequently, to locate
and to characterise the galaxies \citep[e.g.][]{bunker04, bouwens06},
quasars \citep[e.g.][]{fan06} and -- via gamma-ray bursts -- even
individual stars \citep{jakobsson06, haislip06} that existed during
this period of cosmic phase transition, $\sim$0.4--1\,Gyr after the
Big Bang.

\begin{table*}
\caption{Properties of target $z>\rm 5$ quasars and their host galaxies.}
\begin{tabular}{lcrcccccc}
\hline            
Source Name       & $z$  &$M_{\rm B}$ & R.A. & Dec. & $S^{\rm phot}_{\rm 1250\mu m}$ & $S^{\rm phot}_{\rm 850\mu m}$&$S^{\rm phot}_{\rm 450\mu m}$&$S_{\rm 1.4GHz}$ \\
                  &      &      &\multicolumn{2}{c}{(J2000)} &  (mJy)     &  (mJy)&(mJy)    &  ($\mu$Jy)    \\
\hline
SDSS J0756$+$4104 & 5.09 & $-$26.9  & 07 56 18.14 & +41 04 08.6 & 5.5\,$\pm$\,0.5   & 13.3\,$\pm$\,2.1& 14\,$\pm$\,19& 65\,$\pm$\,15 \\
SDSS J1030$+$0524 & 6.28 & $-$27.7  & 10 30 27.10& +05 24 55.0  &
 $<$3.4   & 1.3\,$\pm$\,1.0&$-$21\,$\pm$\,10& $<$61 \\
SDSS J1044$-$0125 & 5.73 & $-$28.0 & 10 44 33.04& $-$01 25 02.1& 2.5\,$\pm$\,0.55 &
6.1\,$\pm$\,1.2& --- &$<$79 \\
\hline
\end{tabular}
\label{tab1}
\end{table*}

Although it is thought that the ultraviolet (UV) light from young
stars, rather than AGN, provided the bulk of the ionizing radiation
\citep{fck06, srbinovsky07}, very high-redshift quasars are
nevertheless important objects in their own right. SMBHs are found
ubiquitously in the cores of nearby galaxies; studying their
accretion-powered growth phase, during which they manifest themselves
as high-redshift AGN, provides a window on the formation and evolution
of the ancestors of present-day massive galaxies. The very existence
of luminous AGN at redshifts above five imposes important constraints on
processes governing the formation of massive galaxies, SMBHs and
proto-clusters in the early Universe. Their bolometric powers imply
accretion onto SMBHs at a prodigious rate
($\sim$1\,M$_{\odot}$yr$^{-1}$, assuming Eddington-limited accretion).
Measurements based on the line width of Mg\,{\sc ii} indicate SMBH
masses of $\sim$10$^{9.5}$\,M$_{\odot}$ \citep{wmj03}, implying a host
dark matter halo of mass $\sim$10$^{12.5}$-M$_{\odot}$
\citep{croom05}.  Submm photometry of the host galaxies of $z>\rm 5$
quasars reveal copious quantities of dust ($\sim10^{8-9}$\,M$_{\odot}$
-- \citealt{priddey03, robson04}) and CO emission lines imply
$\sim10^{10-11}$\,M$_{\odot}$ of molecular gas \citep{walter03,
walter04}. Both are indicative of strong star formation; indeed, a
colossal starburst is required if the dust has been synthesised on the
short timescale available ($\la$1\,Gyr since the Big Bang;
$\la$600\,Myr since the onset of reionization).

The existence of massive, collapsed haloes at high redshift might
appear to conflict with the prevailing hierarchical cold dark matter
(CDM) model of structure formation, in which the initial density
fluctuations have higher power on smaller scales. Using the
Press--Schechter approximation, for example, \citet{er88} predicted a
marked decline in the number density of quasars at $z>\rm 5$. To
accommodate the growing sample of luminous AGN spectroscopically
confirmed to lie at $z>\rm 5$ within the CDM scheme, they are
identified as corresponding to rare, high-$\sigma$ peaks in the
initial overdensity distribution. As such, they constitute excellent
laboratories in which to test the predictions of structure-formation
models operating {\it in extremis}. Specifically, high-$\sigma$ haloes
should be {\it biased} -- more spatially correlated than the
underlying mass -- since neighbouring fluctuations sitting atop a
low-amplitude, large-scale mode of the overdensity spectrum are
boosted over the collapse threshold sooner \citep{kaiser84}. The
vicinity of a high-redshift, luminous AGN would thus be expected to
contain an overabundance of massive, actively star-forming galaxies,
the progenitor of a rich cluster at $z=\rm 0$.  More recently, the
phenomenon of ``downsizing'' provides a framework in which galaxy
formation at high redshift is completed earlier in more massive
objects.  Specifically, the semi-analytic ``Anti-hierarchical Baryon
Collapse'' model of \citet{granato04} suggests an intimate
evolutionary link between quasars and SMGs, feedback from the AGN
playing an important role in regulating star formation in the host.

The submm waveband is an eminently appropriate place in which to
search for structure, as traced by dusty, star-forming galaxies
surrounding high-redshift AGN. This conjecture is supported by
submm/mm imaging campaigns which have identified an excess of SMGs in
fields centred on $z\sim\rm 4$ radio galaxies \citep{ivison00,
stevens03, debreuck04, greve07} and a $z\sim\rm 2$ absorbed QSO
\citep{stevens04}. In this paper, we extend this work to higher
redshift, and expand upon our earlier photometric submm programmes
\citep{isaak02,priddey03a},
obtaining very deep submm imaging of the fields of three
$z>\rm 5$ radio-quiet quasars.

Throughout, we assume cosmological parameters $\Omega_{\rm M}=0.27$,
$\Omega_\Lambda=0.73$ and $H_0=71$\,km\,s$^{-1}$\,Mpc$^{-1}$
\citep{spergel07}.

\section{Observations and Data Analysis}

\subsection{The sample}

Three fields centred on quasars at $z>\rm 5$ were selected for
observation. The quasars were all discovered as part of the Sloan
Digital Sky Survey (SDSS) by \citet{fan00, fan01} and
\citet{anderson01}. Their properties are listed in Table~\ref{tab1}:
all three are luminous at optical wavelengths (absolute $B$ magnitudes
in Table~\ref{tab1} are extrapolated from 1450\AA\ continuum flux
assuming a spectral index, $-$0.5), implying black hole masses of
$\sim$10$^{9.3-9.7}$\,M$_{\odot}$, assuming a bolometric correction
from the $B$ band of 12 \citep{elvis94} and Eddington-limited
accretion. The sample was also designed to provide contrast in
far-infrared luminosity, containing both the brightest member of the
\citet{priddey03} submm photometry sample (SDSS J0756+4104, $S_{\rm
850\mu m}=\rm 13.4$\,mJy) and the deepest upper limit (SDSS
J1030+0524, $S_{\rm 850\mu m}<\rm 2$\,mJy).

\subsection{Observations}
\label{obs}

Jiggle-map observations were made using the SCUBA bolometer array
\citep{holland99} on the 15-m JCMT over the period between 2001 Feb 28
and 2002 Mar 23. In each case the array was centred on the optical
position of the quasar. Every 16\,s the telescope was nodded in either
right ascension or declination by 30\,arcsec in an ON--OFF--OFF--ON
sequence, with the secondary mirror chopping continually at
$\sim$7\,Hz between the ON and OFF positions. For SDSS J1044$-$0125,
we nodded and chopped east-west (E--W) throughout. For SDSS J0756+4104
and SDSS J1030+0524 we nodded and chopped north-south (N--S) as the
target rose and set, and E--W when the target was near transit. By
adopting this latter strategy we ensured the chop direction was as
close to azimuthal as possible and that the region within 30\,arcsec
of the outer bolometers was chopped onto the array; also, using
several position angles reduced the chance of chopping one submm
source systematically onto another.

Dynamic scheduling at JCMT ensured weather conditions were
exceptionally good for our observations. The sky was stable and very
transparent, with a median $\tau^{\rm zenith}_{\rm 225GHz}$ of around
0.043 (often $<$0.04, never worse than 0.09). Flux calibration was
determined by imaging CRL\,618, Mars and Uranus and is accurate to 10
percent at 850\,$\mu$m and 25 percent at 450\,$\mu$m. A long imaging
scan of 3C\,345 was obtained, using the same observing strategy, to
establish an accurate point spread function (PSF).

The total number of integrations/time expended on each field were:
SDSS J0756+4104 (304 integrations; 38.9\,ks); SDSS J1030+0524 (280
integrations; 35.8\,ks); SDSS J1044$-$0125 (312 integrations;
39.9\,ks).

In addition, we publish here for the first time a 250GHz on-off
photometric observation of SDSS J1044$-$0125 obtained using the Max
Planck Institut f\"{u}r Radioastronomie Millimetre BOlometer (MAMBO)
array on the Institute de Radioastronomie Millim\'{e}trique (IRAM)
30-m telescope on Pico Veleta, Spain. The sky opacity during the
observations was in the range 0.164--0.260. Data were reduced using
the {\sc gildas} software package, yielding a 1.2-mm flux density
2.5\,$\pm$\,0.6\,mJy (4.5$\sigma$).

\subsection{Data reduction}

Data from SCUBA were reduced in a standard manner with the {\sc surf}
package \citep{jl98}, using modifications to the {\sc setbolwt} and
{\sc rebin} tasks developed for the survey of high-redshift radio
galaxies by \citet{ivison00} and \citet{stevens03}, as described by
\citet{ivison06}. Briefly, we created an astrometric grid of
1-arcsec$^2$ square pixels then determined accurately weighted and
calibrated signal and noise measurements for each pixel, each
measurement being wholly independent of values for its neighbouring
pixels, with 13.4-arcsec {\sc fwhm} resolution at 850\,$\mu$m. These
signal and noise values reflect the stream of data collected when
bolometers are centred within the region of sky corresponding to a
particular pixel. These unsmoothed images were used for source
detection (see \S\ref{extraction}).

\begin{figure}
\epsfig{figure=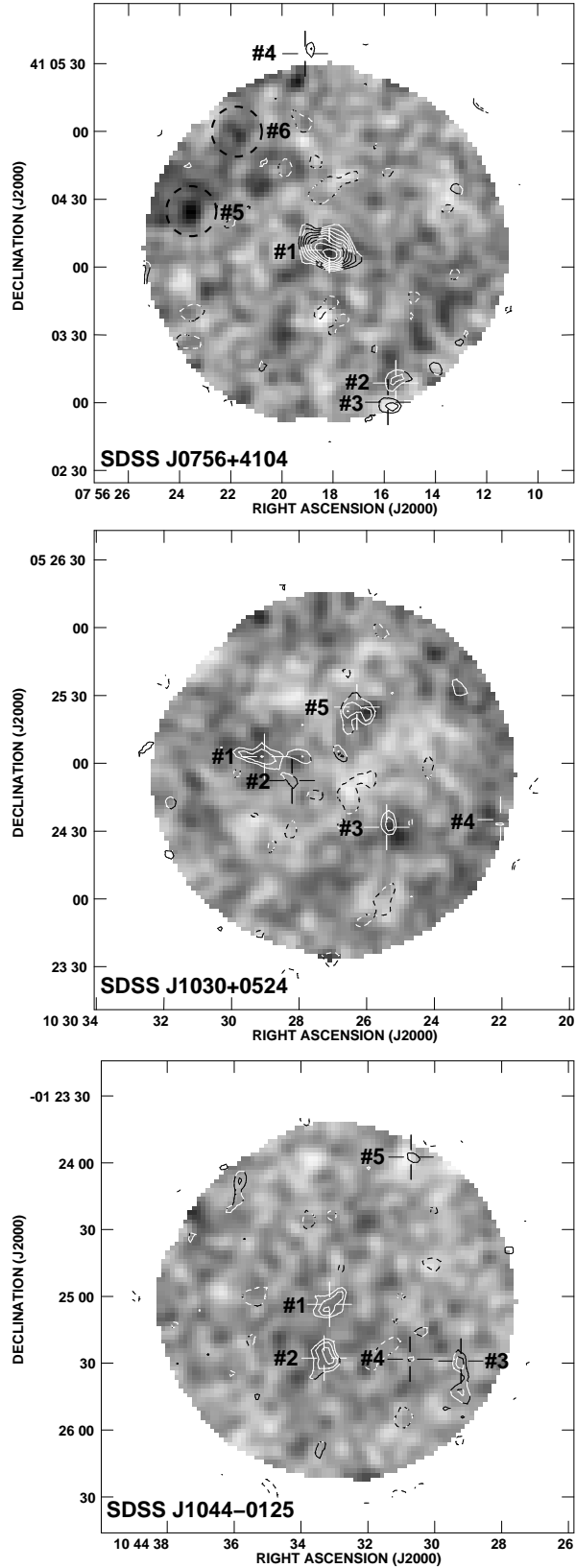,width=79mm}
\caption{submm images of fields centred on three optically selected,
radio-quiet quasars at $z>\rm 5$. The 850-$\mu$m maps are plotted as
contours, at levels -3, +3, 4, 5, 6... $\times \sigma$. The 450-$\mu$m
images are displayed as greyscales. 850-$\mu$m sources (at
$\ge$3$\sigma$ significance) are identified with open crosses; their
labels correspond to those in column 2 of Table~\ref{tab2}. Sources
detected only at 450\,$\mu$m are circled.}
\label{scuba}
\end{figure}

The submm images of each of the three quasar fields -- 850-$\mu$m
contours superposed on greyscale representations of the 450-$\mu$m
data -- are shown in Fig.~\ref{scuba}, where we have smoothed with 6-
and 3-arcsec {\sc fwhm} Gaussians and the resulting 850- and
450-$\mu$m maps have typical noise levels of 1.5 and
7\,mJy\,beam$^{-1}$. For comparison, the 5$\sigma$ 850-$\mu$m
confusion limit for the JCMT is $\approx$1--1.5\,mJy, so we have
reached close to the depth at which confusion is likely to begin
posing problems.

\subsection{Source detection}
\label{extraction}

\begin{table*}
\caption{Properties of SMGs detected in quasar fields.}
\begin{tabular}{lcccccc}
\hline
IAU Source Name	& Id in &R.A.        & Dec.	& $S_{\rm 850\mu m}$  
& S/N & $S_{\rm 450\mu m}$ \\
&Fig.~\ref{scuba}&\multicolumn{2}{c}{(J2000)} &$\pm\sigma$ (mJy)$^{\dagger}$ &       &$\pm\sigma$ (mJy)$^{\dagger}$\\
\hline
SDSS J0756+4104          &\#1 & 07 56 18.15 &  +41 04 07.5  & 13.4\,$\pm$\,1.0 &13.9 &16\,$\pm$\,5 \\
SMM J075615.55+410308.7  &\#2 & 07 56 15.55  & +41 03 08.7  &  4.4\,$\pm$\,1.4 & 3.1 &   3$\sigma < 24$ \\
SMM J075615.86+410300.2  &\#3 & 07 56 15.86  & +41 03 00.2  &  4.6\,$\pm$\,1.4 & 3.1 &   3$\sigma < 24$ \\
SMM J075619.09+410534.4  &\#4 & 07 56 19.09  & +41 05 34.4  & 13.1\,$\pm$\,3.4 & 3.9 &   $\ddagger$        \\
SMM J075621.83+410459.9  &\#5 & 07 56 21.83  & +41 04 59.9  & 3$\sigma < 5.1$ & -- & 56\,$\pm$\,13 \\ 
SMM J075623.58+410424.6  &\#6 & 07 56 23.58  & +41 04 24.6  & 3$\sigma < 5.1$ & -- & 25\,$\pm$\,8 \\
	    		      &              &              &                  &     &        \\
SMM J103029.03+052502.9  &\#1 & 10 30 29.03  & +05 25 02.9  &  7.5\,$\pm$\,1.7 & 4.3 &  42\,$\pm$\,12 \\
SMM J103028.21+052452.3  &\#2 & 10 30 28.21  & +05 24 52.3  &  5.3\,$\pm$\,1.7 & 3.3 &   3$\sigma < 33$\\
SMM J103025.41+052431.7  &\#3 & 10 30 25.41  & +05 24 31.7  &  6.8\,$\pm$\,1.7 & 4.3 &  49\,$\pm$\,11 \\
SMM J103022.04+052435.0  &\#4 & 10 30 22.04  & +05 24 35.0  &  7.0\,$\pm$\,2.2 & 3.3 &   3$\sigma < 48$\\
SMM J103026.29+052524.9  &\#5 & 10 30 26.29  & +05 25 24.9  &  8.5\,$\pm$\,1.7 & 5.0 &   45\,$\pm$\,10 \\
	    		      &              &              &                  &     &        \\
SDSS J1044$-$0125          &\#1& 10 44 33.15 &$-$01 25 03.6 &  5.6\,$\pm$\,1.0 & 6.2 &   3$\sigma < 24$\\
SMM J104433.31$-$012527.7  &\#2& 10 44 33.31 &$-$01 25 27.7 &  7.5\,$\pm$\,1.5 & 5.5 &  24\,$\pm$\,7 \\
SMM J104429.22$-$012529.0  &\#3& 10 44 29.22 &$-$01 25 29.0 &  6.1\,$\pm$\,1.7 & 3.5 &   3$\sigma < 24$\\
SMM J104430.74$-$012528.2  &\#4& 10 44 30.74 &$-$01 25 28.2 &  7.0\,$\pm$\,1.7 & 4.5 &   3$\sigma < 24$\\
SMM J104430.71$-$012357.5  &\#5& 10 44 30.71 &$-$01 23 57.5 &  6.8\,$\pm$\,1.9 & 3.0 &   $\ddagger$\\
\hline
\end{tabular}

\noindent
$\dagger$ Errors exclude the uncertainty in absolute flux calibration:
10 (25) per cent at 850\,$\mu$m (450\,$\mu$m).\\
$\ddagger$ Outside region covered by 450-$\mu$m data.

\label{tab2}
\end{table*}

Source detection at 850\,$\mu$m was accomplished using the algorithm
described in detail by \citet{scott02}, utilising the signal and noise
maps and a PSF (measured for the blazar, 3C\,345) to perform a
simultaneous maximum-likelihood fit to all the potentially significant
peaks in each map. Sources down to a significance level of 3\,$\sigma$
are listed in Table~\ref{tab2}.

Adopting a 3$\sigma$ detection threshold, at 850\,$\mu$m (450\,$\mu$m)
a total of 14 (7) sources are detected: 2 (1) QSOs and 12 (6)
companions in the three fields, with flux densities between 4.4 (16)
and 13.4\,mJy (56\,mJy). Five sources, including one QSO, are detected
at both wavelengths (Table~\ref{tab2}).

\begin{figure}
\begin{center}
\epsfig{figure=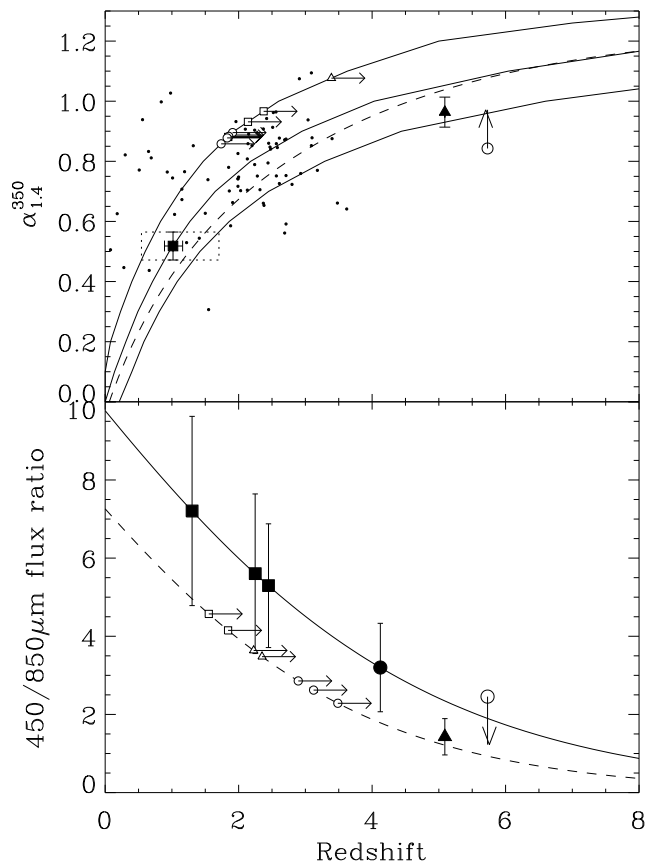,width=85mm}
\end{center}
\caption{{\bf Top:} submm (850\,$\mu$m) to radio (1.4\,GHz) spectral
index as a redshift indicator. The curves show empirical calibrations
from (thick solid, with $\pm$1$\sigma$ shown as thin solid)
\citet{cy00} and (dashed) \citet{yc02}. Also for comparison are
plotted (small points) SMGs with spectroscopic redshifts and radio
fluxes from \citet{chapman05}.  Sources in the SDSS J0756+4104 field
are shown as triangles; in the SDSS J1030+0524 field, squares; in the
SDSS J1044$-$0125 field, circles.  Radio detections have filled
symbols; upper limits are unfilled.  Two of the quasars (SDSS
J0756+4104 and SDSS J1044$-$0125) are plotted at their spectroscopic
redshifts. A tentative radio counterpart to an SMG in the SDSS
J1030+0524 field (\#5) is shown with a dotted box delineating the
1$\sigma$ uncertainty in its photometric redshift.  Radio-undetected
SMGs are plotted at the most conservative lower limit on their
redshift, given by the upper curve from \citet{cy00}.  {\bf Bottom:}
450\,$\mu$m/850\,$\mu$m ratio as a redshift indicator.  The curves are
derived from fits to the spectral energy distributions (SEDs) of
$z>\rm 4$ quasars (\citealt{pm01}, solid) and SMGs
(\citealt{coppin07}, dashed).}
\label{cy}
\end{figure}

Discussing each of the fields in turn:

\begin{description}
\item[{\bf SDSS J0756+4104}] ($z$ = 5.09) was detected via photometry-mode
(on-off) observations at 850\,$\mu$m: 13.4\,$\pm$\,2.1\,mJy
\citep{priddey03} and at 1200\,$\mu$m: 5.5\,$\pm$\,0.5\,mJy
\citep{petric03}. Submm emission is clearly visible in the maps
(Fig.~\ref{scuba}) at the optical coordinates of the quasar (the map
centre), at both 850\,$\mu$m (11.2\,$\pm$\,1.0\,mJy) and 450\,$\mu$m
(16\,$\pm$\,5\,mJy). To the eye, the source appears elongated; the
best-fit 2-D Gaussian corresponds to (21.4\,$\pm$\,9.0) arcsec
$\times$ (13.6\,$\pm$\,3.8) arcsec at a position angle (PA) of
69\,$\pm$\,37$^{\circ}$.

Three further sources were detected ($>$3$\sigma$) in the SDSS
J0756+4104 field at 850\,$\mu$m, each close the edge of the map (one
beyond the areal coverage at 450\,$\mu$m). In addition, two sources
were detected at 450\,$\mu$m, though neither has a robust 850-$\mu$m
counterpart (3$\sigma<\rm 4.5$\,mJy). This casts some doubt on their
reality; certainly, if they are real emitters, they are likely to be
at a {\em much} lower redshift than the central quasar.

\item[{\bf SDSS J1030$+$0524}] ($z$ = 6.28) has been the object of
considerable observational study. \citet{stiavelli05}, for example,
detected two candidate $z\sim\rm 6$ sources in {\it Hubble Space
Telecope (HST)}--ACS images of the field, which they speculate may
belong to the same dark matter halo. The quasar itself
appears similar in all respects to low-redshift AGN: submm photometry
failed to detect the quasar to a sensitive limit of $\rm
3\sigma<3.0$\,mJy at 850\,$\mu$m \citep{priddey03} and a deep X-ray
spectrum taken with {\it XMM-Newton} \citep{farrah04} finds a photon
index $\Gamma\approx\rm 2.1$ with no evidence for intrinsic absorption
($N_{\rm H}<\rm 8\times 10^{22}$\,cm$^{-2}$). The lack of gas and
dust, plus the near-canonical X-ray spectral slope, suggest that this
SMBH and its host have already passed through their major formation
phase, only $\sim$900\,Myr after the Big Bang.

Our submm images (Fig.~\ref{scuba}) confirm the submm quiescence of
the quasar/host galaxy, with 3-$\sigma$ limits of 5.1 and 30\,mJy at
850 and 450\,$\mu$m, respectively. Indeed, there is an interesting
negative feature just south-west of the quasar corresponding to the
western ``off'' position of the weakest SMG in this field, probably
the primary reason SMM J103028.21+052452.3 achieves an overall
significance of $\ge$3$\sigma$.

The surrounding field is crowded: five sources at 850\,$\mu$m and
three at 450\,$\mu$m. Each of the 450-$\mu$m sources lies within a
fraction of a beamwidth of significant 850-$\mu$m emission, confirming
the reality of both.

Is it surprising that the least active quasar host should have the
most active environment?  It is possible that, by targeting luminous,
{\it optically}-selected quasars, we are selecting very massive
objects that are already close to the end of their star-forming
phase. By the rules of ``downsizing'', the evolution of the most
massive objects is completed earliest, so it should not be surprising
to see companions to fully-formed, high-redshift quasars in an earlier
evolutionary stage (see also Section~\ref{model}).

A 1.4-GHz Very Large Array image ($\sigma\approx\rm 20\,\mu$Jy) of the
field was obtained by \citep{petric03}, comparison with which enables
us to pinpoint counterparts at other wavelengths with sub-arcsec
precision \citep[e.g.][]{ivison98}. In addition, constraints can be
placed on the redshifts of those sources with radio counterparts, or
meaningful limits. SMM J103029.03+0452502.9 lies 6.8\,arcsec from a
radio source at $\alpha\rm = 10^h 30^m 26.^s0, \delta = +05^{\circ}
25' 19.''69$ with an integrated 1.4-GHz flux density of
427\,$\pm$\,52\,$\mu$Jy (Andreea Petric, private communication), giving
a spectral index, $\alpha^{\rm 350GHz}_{\rm 1.4GHz}$ =
0.52\,$\pm$\,0.04. If the submm and radio sources are the same (and
this is statistically very probable -- $P<\rm 0.05$ -- see
\citealt{ivison02}), the FIR--radio correlation for star-forming
galaxies predicts $z=\rm 1.0^{+0.7}_{-0.5}$ (\citealt{cy00}; see
Figure~\ref{cy}).  SMM J103025.41+052431.7 is near-coincident
(separation, 4.4\,arcsec) with the core of a limb-brightened
Fanaroff--Riley {\sc ii} (FR\,{\sc ii}) radio galaxy (Fig.~2 of
\citeauthor{petric03}) -- another statistically significant
association, i.e.\ $P<\rm 0.05$. This raises the strong suspicion,
allied with their relatively high 450-$\mu$m flux densities, that some
or all of the other nearby SMGs might lie in the foreground,
associated with this radio galaxy, rather than at the redshift of the
quasar. Such luminous radio galaxies are commonly found in clusters
\citep{best03, best07} and some clusters are known to host SMGs
\citep{best02, webb05}.

\item[{\bf SDSS J1044$-$0125}] ($z$ = 5.73) has an optical spectrum with a
C\,{\sc iv} absorption trough suggesting that this is a broad
absorption line (BAL) quasar \citep{maiolino01, goodrich01}.
Correspondingly, the quasar is weak in X-rays \citep{brandt01},
consistent with the BAL interpretation \citep[e.g.][]{gallagher06}.
The quasar/host was detected at 850 and 1200\,$\mu$m using on-off
measurements with SCUBA and MAMBO respectively: 6.1\,$\pm$\,1.2\,mJy
and 2.5\,$\pm$\,0.6\,mJy (see Section~\ref{obs}).

In our SCUBA image, the quasar is detected robustly at 850\,$\mu$m
with a flux density of 5.7\,$\pm$\,1.5\,mJy, with weak evidence of
extended emission (Section~\ref{extended}). Four further sources are
detected at 850\,$\mu$m; one at 450$\mu$m, corresponding to the
brightest 850-$\mu$m emitter.
\end{description}

\section{Discussion and analysis}

\subsection{Photometric redshifts}
\label{sec:photoz}

We can place crude constraints on the redshift of any 850-$\mu$m
emitters in our quasar fields using detections or limits at 1.4\,GHz
and/or 450\,$\mu$m. Much of the uncertainty in this procedure rests in
the intrinsic diversity of SEDs exhibited by SMGs \citep{ivison-div}.
Radio imaging at 1.4\,GHz of the three fields, reaching rms levels
of 17, 20 and 27\,$\mu$Jy beam$^{-1}$ for SDSS J0756+4104, SDSS
J1030+0524 and SDSS J1044$-$0125, respectively, were published by
\citet{petric03}. As noted earlier, two sources in the SDSS J1030+0524
field have radio counterparts: identification of SMM
J103026.29+052524.9 with a radio emitter implies $z\sim\rm 1$, and SMM
J103025.41+052431.7 is coincident with the core of a FR\,{\sc ii}
radio galaxy. Of the two quasars with submm detections, SDSS
J0756+4104 has a 1.4-GHz flux density of $65\pm17\,\mu$Jy, implying a
spectral index, $\alpha^{\rm 350GHz}_{\rm 1.4GHz}=0.93\pm0.05$; SDSS
J1044$-$0125 has a radio upper limit, $3\sigma<\rm 80\,\mu$Jy, giving
$\alpha^{\rm 350GHz}_{\rm 1.4GHz}>\rm 0.85$ (2$\sigma$). Both are
consistent with starbursts, given their respective redshifts
(Fig.~\ref{cy}). In Fig.~\ref{cy} we also indicate the lower limits on
$\alpha^{\rm 350GHz}_{\rm 1.4GHz}$ corresponding to SMGs {\it not}
detected in the \citet{petric03} radio maps: all are consistent with
the sources lying at high redshift, the weakest lower limits implying
$z>\rm 1.5$ (assuming a generous scatter in $\alpha^{\rm 350GHz}_{\rm
1.4GHz}$.)

Fig.~\ref{cy} also illustrates the 450\,$\mu$m/850\,$\mu$m flux
density ratio as a redshift indicator assuming idealised, isothermal,
greybody dust spectra: one SED determined from $z>\rm 4$ quasars by
\citet{pm01} with $T_{\rm d}=42$\,K, $\beta=\rm 1.9$; another SED
determined from blank-field SMGs in the SCUBA HAlf Degree
Extragalactic Survey (SHADES) with $T_{\rm
d}=35$\,K, $\beta=\rm 1.5$ \citep{coppin07}.
A low 450\,$\mu$m/850\,$\mu$m flux density
ratio implies a high redshift since the observed 450-$\mu$m emission
must have originated at the peak, or possibly on the Wien side, of the
SED. Thus it is plausible that any 850-$\mu$m sources undetected in
deep 450-$\mu$m data lie at high redshift, the most conservative
constraints giving $z\ga\rm 2$.  Conversely, a 450-$\mu$m emitter with
no corresponding 850-$\mu$m detection can be considered to have a weak
upper limit on its redshift.

Ideally, one would search for spectral features to establish the
redshifts of the companion sources unambiguously.  The rest-frame
mid-IR, FIR and submm wavebands offer ionic/atomic fine-structure
lines as well as the rotational transitions of molecular CO and
HCN. Unfortunately, such observations are beyond the reach of current
ground- and space-based instrumentation, though this will change with
the advent of the Atacama Large Millimetre Array (ALMA) and future
actively cooled space missions such as the Space Infrared Telescope
for Cosmology and Astrophysics (SPICA -- \citealt{nakagawa04}) or the
FIR Interferometer (FIRI -- \citealt{hi07}).

\subsection{Number counts}
\label{sec:counts}

Earlier submm/mm imaging surveys have shown evidence for
overdensities of SMGs in the vicinities of AGN at high redshift: in
fields around radio galaxies at $z\sim\rm 4$ \citep[e.g.][]{ivison00,
stevens03, debreuck04, greve07} and in the field of an absorbed QSO at
$z\sim\rm 2$ \citep{stevens04}.  Are our $z>\rm 5$ QSO fields
consistent with these findings?

\begin{figure}
\epsfig{figure=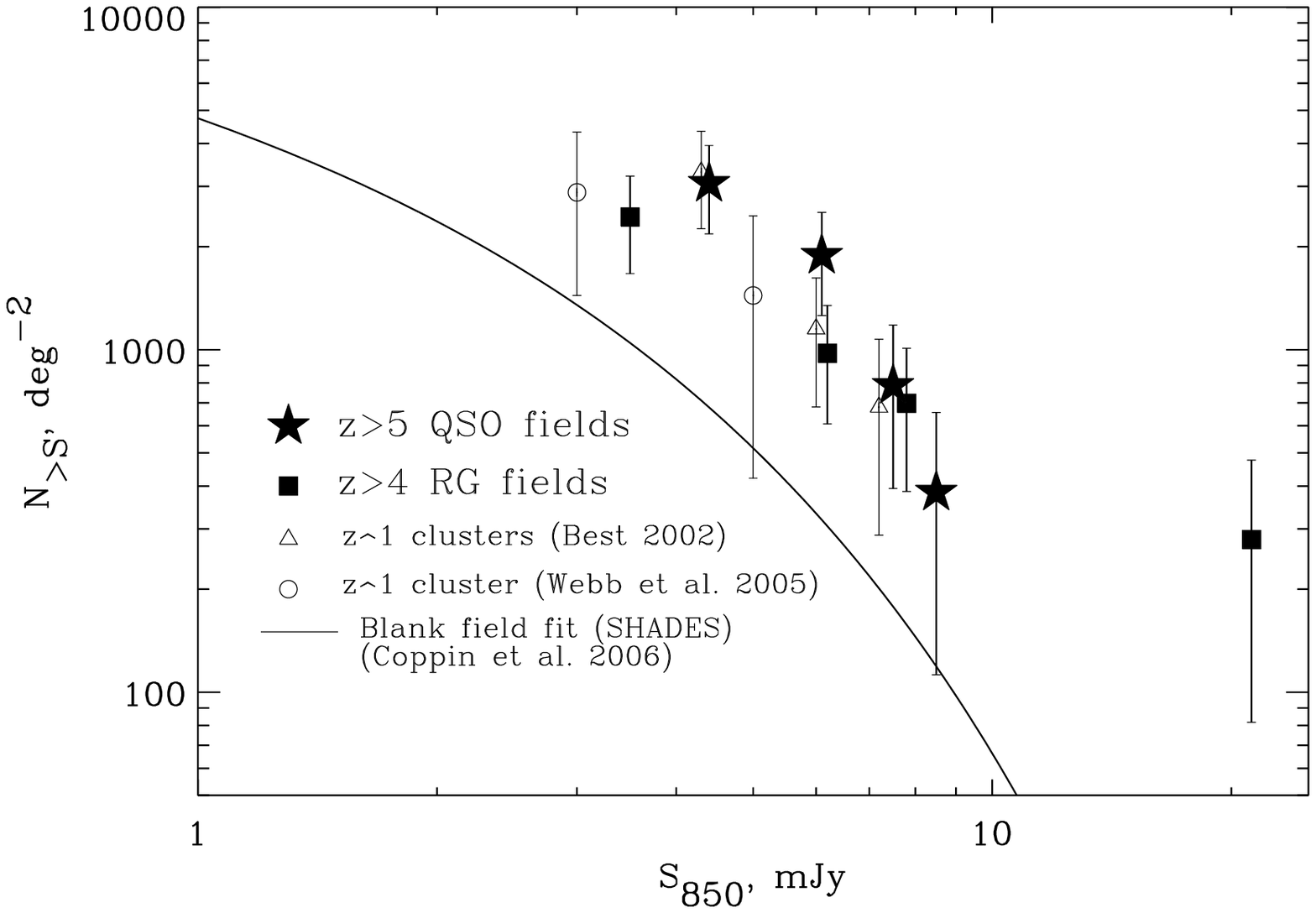,width=85mm}
\caption{Cumulative number counts of SMGs, in fields centred on (i)
luminous high-redshift AGN (filled symbols): $z>\rm 5$ optically
selected quasars (stars; this work); $z\sim\rm 4$ radio galaxies
(squares; \citealt{stevens03}); and (ii) $z\sim\rm 1$ clusters
(unfilled symbols -- \citealt{best02, webb05}). The line shows a
functional fit to the blank-field number counts of SMGs in the SHADES
submm survey \citep{coppin06}. Points have been corrected for
effective area as a function of survey depth, but not for flux
boosting, confusion or incompleteness. Even so, the excess over the
blank-field counts is striking.}
\label{counts}
\end{figure}

Fig.~\ref{counts} compares the cumulative number counts observed in
our three $z>\rm 5$ quasar fields (``z5Q'') with (a) counts from
\citealt{stevens03} (``HzRG''); (b) blank-field submm survey counts (a
fit to the SHADES counts by \citealt{coppin06}); (c) counts from submm
imaging surveys of clusters at $z\sim\rm 1$ \citep{best02, webb05}.
The raw number counts have been corrected for effective survey area as
a function of limiting flux density. However, we have not corrected
for incompleteness or for flux boosting. At the faintest levels, or in
noisy regions of the map where sources lie near the flux limit, these
effects are significant, but they are difficult to correct in maps of
this size and involve assumptions about the number count distribution,
e.g.\ using the blank-field counts with which we are attempting to
compare.

Nevertheless, assuming Poisson errors there is a significant excess
over the blank-field counts. The seven radio galaxy fields and the
three z5Q fields each contain a total of 12 850-$\mu$m companion
sources. The z5Q counts (and those of the $z\sim\rm 1$ clusters) are
consistent with the radio galaxy field counts (with the caveat that
the z5Q sample spans a more limited range in flux density, due to
shallower limit and smaller total area); both contain more SMGs than
blank fields by a factor 4--5 across a range of 850-$\mu$m flux
densities, including the $\gs$6-mJy regime where the effects of flux
boosting are minimal and overall sample reliability is excellent.

In a blank-field survey of equivalent area and depth to all three of
our fields combined, one would expect (based upon model fits to the
SHADES counts from \citealt{coppin06}) to detect $\sim$3, $\sim$2 and
$\sim$1 sources brighter than $S_{850\rm \mu m}$ = 4.1, 5.4 and
7.3\,mJy, respectively. It is thus likely that some of the 12 z5Q
companion sources are foreground contaminants, as we have seen already
in \S\ref{sec:photoz}. Others, as we have also seen, are consistent
with SMGs at high redshift, based on their radio/submm and
450\,$\mu$m/850\,$\mu$m flux density ratios. We conclude that the maps
probably contain SMGs genuinely associated with the dark matter haloes
inhabited by our target quasars, leading to the statistical
overdensities observed.

As well as the potential for foreground objects to
contaminate the sample directly via their submm emission, we should consider
whether the observed excess of SMGs may be due to their 
gravitational lensing effects \citep[e.g.][]{chapman02}.  Depending on the
slope of the luminosity function, \citet{wl02} estimate that
$\sim\frac{1}{14}-\frac{1}{3}$ of $z\sim\rm 6$ quasars could be
lensed.  Targeting two submm-bright quasar hosts may have introduced
an additional bias toward lensed fields, while the possibility that
the submm-faint host is lensed is enhanced by the presence of a
foreground radio galaxy (and possibly a cluster associated with that
radio galaxy). The degree to which number counts are boosted by
lensing depends on a competition between (i) the abundance of faint
sources available to be boosted by the gravitational magnification and
(ii) the stretching of area in the source plane. Since the submm source
counts are steep, (i) is a strong effect. Deep optical images of
$z>\rm 5$ quasars, including SDSS J1030+0524, have failed to reveal
any morphological signatures of strong lensing
\citep{fan03}. Neither are line-of-sight galaxies, that could provide
magnifications greater than $\sim$1.1, in evidence
\citep{willott05}. For now, therefore, we consider the observed
overdensity to be due to high-redshift SMGs in the vicinity of the
signpost quasars, rather than lensing or unusual foreground activity.

\subsection{Star-formation rates and dust masses at $z>\rm 5$}

We can estimate SFRs for SMGs from their 850-$\mu$m flux densities,
$S_{\rm 850\mu m}$, assuming that they lie at the redshifts of the
quasars. The range of FIR luminosity of a source at the average
redshift of the three quasars ($z=\rm 5.7$) is $L_{\rm FIR}$ =
0.7$\rightarrow$1.4 $\times(S_{\rm 850\mu m}/{\rm mJy})\times
10^{12}$\,L$_{\odot}$, assuming thermal SEDs with $T_{\rm d}$ =
35--42\,K and $\beta$ = 1.5--1.9. If $L_{\rm FIR}$ is powered by
reprocessed stellar light, this translates into an instantaneous
formation rate of massive stars of $\Psi\times(L_{\rm FIR}/10^{10}
{\rm L}_{\odot}$)\,M$_{\odot}$\,yr$^{-1}$. Here, $\Psi$ depends upon
factors such as the stellar mass function and the efficiency with
which starlight is reprocessed by dust; we assume $\Psi\sim\rm 1$
giving a SFR range of 70$\rightarrow$140 $\times (S_{\rm 850\mu
m}/{\rm mJy})$\,M$_{\odot}$\,yr$^{-1}$. Summing up the contributions
of the companion SMGs (but not of the quasars, to avoid AGN
contamination, even though it is likely that a substantial fraction of
their submm luminosity is due to star formation) gives a total SFR per
field, averaged over the three maps, of $\approx f_{z>\rm 5}\times$
2000$\rightarrow$4000\,M$_{\odot}$\,yr$^{-1}$, where $f_{z>\rm 5}$
represents the fraction of sources at the quasar redshifts. This star
formation is apparently taking place in a region of volume
$\ls$1\,Mpc$^3$.

Adopting a dust opacity $\kappa_{\rm 125\mu m}=30$\,cm$^2$\,g$^{-1}$
(see \citealt{priddey03} for an explanation), the dust mass
corresponding to $S_{\rm 850\mu m}$ is $M_{\rm d}$ =
0.5$\rightarrow$1.0 $\times (S_{\rm 850\mu m}/{\rm mJy})\times
10^{8}$\,M$_{\odot}$, giving a total dust mass per field of $\approx
f_{z>\rm 5}\times$ 1.6$\rightarrow$2.8 $\times 10^9$\,M$_{\odot}$. In
this case it is valid to include the quasar contribution, regardless
of whether their submm luminosity is powered by starburst or AGN. The
result is $\sim$1.9$\rightarrow$3.3 $\times 10^9$\,M$_{\odot}$). If
the dust-to-gas ratio in these objects is similar to that of the Milky
Way, their total gas mass is $\sim(S_{\rm 850\mu m}/{\rm mJy})\times
10^{10}$\,M$_{\odot}$.

Finally, we consider the possibility that some of the submm companions
are powered by buried AGN rather than obscured star formation.  This
is not implausible given evolutionary scenarios for SMGs involving
co-evolution between black holes and massive spheroids. Assuming, in
the most extreme case, that $L_{\rm FIR}$ represents the bolometric
luminosity of the AGN, and assuming Eddington-limited accretion, the
brightest SMGs ($\approx$10\,mJy) would have black holes of mass
$\approx 4\times 10^8$\,M$_{\odot}$.  Their hard X-ray fluxes
(2--10\,keV) would be $\approx 4\times
10^{-15}$\,erg\,cm$^{-2}$\,s$^{-1}$. For the general SMG population,
or at least the radio-identified subset, deep X-ray observations
\citep[e.g.][]{alexander05} show that their AGN are weak compared with
their bolometric luminosity. We have no reason to believe the SMGs in
our high-redshift quasar fields should be any different; indeed, their
SMBHs are likely somewhat less developed than those explored by
\citeauthor{alexander05} at $z\sim\rm 2.2$. 

\subsection{Resolved emission and radial profiles}
\label{extended}

To measure source sizes requires knowledge of the PSF, determined to
be 13.4 arcsec {\sc fwhm} and near-circular by fitting a 2-dimensional
(2-D) Gaussian to a beam map of 3C\,345. We have used these parameters
to determine the size of the most significant ($>6\sigma$) sources,
again using 2-D Gaussian fits within the {\sc aips} software
environment.

SDSS J0756+4104 appears to be resolved along an axis with PA
69$^{\circ}$. Along the orthogonal axis, the quasar's submm emission is
point-like. Deconvolving the beam from the best-bet 2-D Gaussian fit
suggests an emission region of size $\rm (16.0\pm 1.5)$\,arcsec
$\times$ $\rm (0.0\pm 1.5)$\,arcsec, though the apparent morphology
could be mimicked by two or more well-separated, compact sources.

At first sight SDSS J1044$-$0125 also appears resolved, albeit with
less certainty than for the $\sim$14$\sigma$ detection of SDSS
J0756+4104.  However, the best-fit 2-D Gaussian has dimensions similar
to those of the beam: $\rm (14.4\pm 2.5)$\,arcsec $\times$ $\rm
(12.5\pm 2.5)$\,arcsec at PA 125$^{\circ}$, so we conclude that there
is no compelling evidence that the emission from SDSS J1044$-$0125 has
been resolved.

The angular scale is $\approx$6\,kpc\,arcsec$^{-1}$ at $z\sim\rm 5-6$,
so the physical scale of the emission from SDSS J0756+4104 corresponds
to $\approx$100\,kpc on the plane of the sky. This suggests, perhaps,
that we are witnessing a colossal merger. Lensing might provide an
alternative explanation, though there is scant evidence for this from
optical images.

Evidence of source structure deviating from that of a point source has
been noted in previous submm observations of high-redshift galaxies
\citep[e.g.][]{ivison00, stevens03}. The mm continuum and CO line
emission from the $z=\rm 4.7$ quasar, BR 1202$-$0725, was clearly
separated into two components by \citet{omont96}, albeit with a
smaller angular separation (4\,arcsec) than single-dish submm imaging
is capable of resolving. Several high-redshift radio galaxies have
also been resolved in CO line emission \citep{pppp00, debreuck05} on
arcsec scales using the IRAM Plateau de Bure interferometer (PdBI).

\subsection{Comparison with model predictions}
\label{model}

The linear Press--Schechter approximation can be used to demonstrate
analytically the concept of bias in a hierarchical cosmology
\citep[e.g.][]{mo+white02}.  However, numerical simulations of the
evolution of CDM, which can track the collapse of fluctuations on
large scales into the non-linear regime, reveal a more complex
picture, the dark matter exhibiting a rich spatial structure
(filaments, clusters: the so-called ``cosmic web'').

In the ``Millenium Simulation'', \citet{springel05} specifically
identify $z=\rm 6$ quasars, selecting them as the objects with the
most massive dark matter haloes and/or the largest stellar mass.
These objects have halo masses $\sim10^{12.5}$\,M$_{\odot}$, SFRs of
several 100\,M$_{\odot}$\,yr$^{-1}$ and evolve into the central
dominant galaxies of rich clusters at $z=\rm 0$. At $z=\rm 6$ they are
surrounded by numerous star-forming galaxies and lie on prominent dark
matter filaments (Fig.~3 of \citealt{springel05}). The knots strung
out on these filaments are reminiscent of the submm companions to
$z>\rm 5$ quasars -- especially, for example, the neighbouring pair in
the SDSS J0756+4104 image (Fig.~\ref{scuba}), and the possible
binarity of the quasar host itself.  However, such claims are
premature: the angular scale over which one would confidently expect
to see large-scale structure is somewhat greater than that enclosed
within the field-of-view of SCUBA. The present images correspond only
to the innermost $\sim$1\,Mpc of such structures.

A tentative evolutionary scheme in which to interpret our observations
is the ``anti-hierarchical'' model of \citet{granato04, granato06}.
This scheme not only incorporates the effects of feedback from the AGN
on star formation, but explicitly takes account of dust.  A
counter-intuitive feature of this model is that the evolution
progresses more rapidly for the most massive objects. SMGs are
envisaged as massive spheroids undergoing a major episode of
dust-enshrouded star formation, containing small (but gradually
accreting) black holes in their cores. Eventually the black hole
becomes sufficiently massive to power a quasar, thereby terminating
star formation via jets in radio-loud systems or via accretion-driven
winds in radio-quiet objects. The system then evolves passively as a
massive elliptical galaxy to the present day. Submm-luminous
high-redshift quasars presumably correspond to the late stages of this
transition between SMG and QSO \citep{page04, stevens05}. If the QSO
represents the most massive collapsed object in its field (by design
we have selected extremely luminous objects), it will have a head
start over the SMGs in its field. As suggested earlier, it may not be
surprising that SDSS J1030+0524 is ``submm dead'', despite residing in
the most active field: it may have evolved more rapidly than its
companions and expended its fuel.

\section{Summary}

Submm images at 450 and 850\,$\mu$m of the fields of optically
luminous $z>\rm 5$ quasars reveal an excess of submm emitters relative
to expected counts of SMGs in blank fields.  Although flux ratios show
that some of the companion SMGs are undoubtedly foreground objects,
the submm counts suggest that many of these companions probably lie in
the dark matter haloes inhabited by our target quasars

Two of the quasars are luminous submm emitters in their own right,
suggesting that their hosts are intensely star-forming
galaxies. Furthermore, one shows tentative signs of extended submm
emission, an indication either that the host galaxy is resolved (on a
$\sim$100-kpc scale) or, more likely, is caught during a merger. This
finding reinforces, and extends to higher redshift, the conclusions of
previous submm/mm imaging surveys of high-redshift AGN.

The observed overdensities of luminous, star-forming galaxies are
consistent with the idea that luminous AGN reside in the most massive
dark matter haloes at any epoch and, as such, pinpoint highly biased
regions in the high-redshift Universe which eventually merge to become
rich clusters in the present day. Any more detailed interpretation is
hindered by the limitations of present instrumentation but forthcoming
FIR and submm facilities will enable significant progress to be made.

\begin{itemize}

\item At current sensitivity levels, set by confusion at $\sim$2\,mJy
at 850\,$\mu$m, we are sensitive only to the very brightest starbursts
-- several $\times 100$\,M$_{\odot}$\,yr$^{-1}$ of star formation, not
at all typical of known $z>5$ field galaxies.  However in the future,
facilities such as ALMA will provide a powerful means of locating more
typical star-forming galaxies at high redshift. It is important to
characterise the feasibility of such a project by learning what we can
about the role of dust in the formation of the earliest galaxies. The
presence of dust at these redshifts may have significant implications
for reionization, as well as for Lyman-$\alpha$ line searches
for high-redshift galaxies -- the escape into the intergalactic medium
of UV photons being hindered and thermally degraded by dust grains.

\item The large beam size of the JCMT at 850$\mu$m precludes scrutiny of
the sources on fine spatial scales, yet provides a tantalising
glimpse of partially resolved or interacting galaxies. The enhanced
resolution of submm interferometers such as IRAM PdBI and, ultimately,
ALMA, will allow for more intricate study of the morphology and
dynamics of high-redshift starbursts.

\item The high areal mapping speed offered by SCUBA-2
\citep{holland06} opens up the possibility that we will be able to
trace the distribution of star-forming galaxies on wide scales
surrounding luminous, high-redshift AGN, obtaining surer statistics,
perhaps even tracing the dark matter filaments and thus directly
testing some of the predictions of numerical simulations of CDM
models.

\item Wide-field spectral imaging at mid-IR--mm wavelengths offers the
best prospects with which to determine the redshifts of SMGs in the
quasar fields unambiguously. Whilst such observations are beyond the
capabilities of current facilities, the sensitivities of ALMA and
future FIR space missions will be sufficient to pinpoint dusty,
star-forming galaxies out to $z\sim\rm 6$.

\end{itemize}

The prospects implied by these discoveries for the next generation of
submm instruments are therefore extremely promising.

\section*{Acknowledgements}
RSP gratefully acknowledges support from the University of
Hertfordshire.  The JCMT is operated on behalf of the UK Science and
Technology Facilities Council, the Netherlands Organisation for
Scientific Research and the National Research Council of Canada.

\bibliographystyle{mn2e}
\bibliography{refs}

\end{document}